\documentclass[conference]{IEEEtran}
\IEEEoverridecommandlockouts
\usepackage{cite}
\usepackage{amsmath,amssymb,amsfonts,amsthm}
\usepackage{graphicx}
\usepackage{textcomp}
\def\BibTeX{{\rm B\kern-.05em{\sc i\kern-.025em b}\kern-.08em
    T\kern-.1667em\lower.7ex\hbox{E}\kern-.125emX}}

\usepackage{lettrine}
\usepackage{comment}
\usepackage{enumitem}
\usepackage{booktabs}
\usepackage[table]{xcolor}
\usepackage{multirow}
\usepackage{tablefootnote}
\usepackage{bbm}
\usepackage{bm}
\usepackage{chngpage}
\usepackage{textcomp}
\usepackage{hyperref}
\usepackage{url}
\usepackage[normalem]{ulem}
\usepackage{algpseudocode}
\newsavebox{\ieeealgbox}

\usepackage{array}

\newtheorem{theorem}{Theorem}

\newtheorem{lemma}{Lemma}
\newtheorem{definition}{Definition}

\newtheorem{axiom}{Axiom}
\newtheorem*{policy*}{Dynamic NEM}

\newcommand*{\QED}{\hfill\ensuremath{\square}}

 \def\old#1{}

\def\nn{\nonumber}
\def\beq{\begin{equation}}
\def\eeq{\end{equation}}
\def\bea{\begin{eqnarray}}
\def\eea{\end{eqnarray}}
\def\ba{\begin{array}}
\def\ea{\end{array}}

\def\bitem{\begin{itemize}}
\def\eitem{\end{itemize}}
\def\ben{\begin{enumerate}}
\def\een{\end{enumerate}}

\definecolor{bgrd}{rgb}{1,1,1}
\definecolor{gray}{rgb}{0.5,0.5,0.5}
\definecolor{dkr}{rgb}{0.7,0.1,0.2}
\definecolor{dkb}{rgb}{0.1,0.1,0.8}

\begin{document}

\title{Achieving Social Optimality for Energy Communities via Dynamic NEM Pricing
\thanks{This work was supported in part by the National Science Foundation under Awards 1932501 and 2218110.}
}

\author{\IEEEauthorblockN{Ahmed S. Alahmed and Lang Tong (\{asa278, lt35\}@cornell.edu)}
\IEEEauthorblockA{\textit{School of Electrical and Computer Engineering, Cornell University}, Ithaca, USA}
}

\maketitle

\begin{abstract}
We propose a social welfare maximizing mechanism for an energy community that aggregates individual and shared community resources under a general net energy metering (NEM) policy. Referred to as Dynamic NEM, the proposed mechanism adopts the standard NEM tariff model and sets NEM prices dynamically based on the total shared renewables within the community. We show that Dynamic NEM guarantees a higher benefit to each community member than possible outside the community. We further show that Dynamic NEM aligns the individual member's incentive with that of the overall community; each member optimizing individual surplus under Dynamic NEM results in maximum community's social welfare. Dynamic NEM is also shown to satisfy the cost-causation principle. Empirical studies using real data on a hypothetical energy community demonstrate the benefits to community members and grid operators.
\end{abstract}


\section{Introduction}\label{sec:intro}

\begin{figure}[htbp]
    \centering
    \includegraphics[scale=0.45]{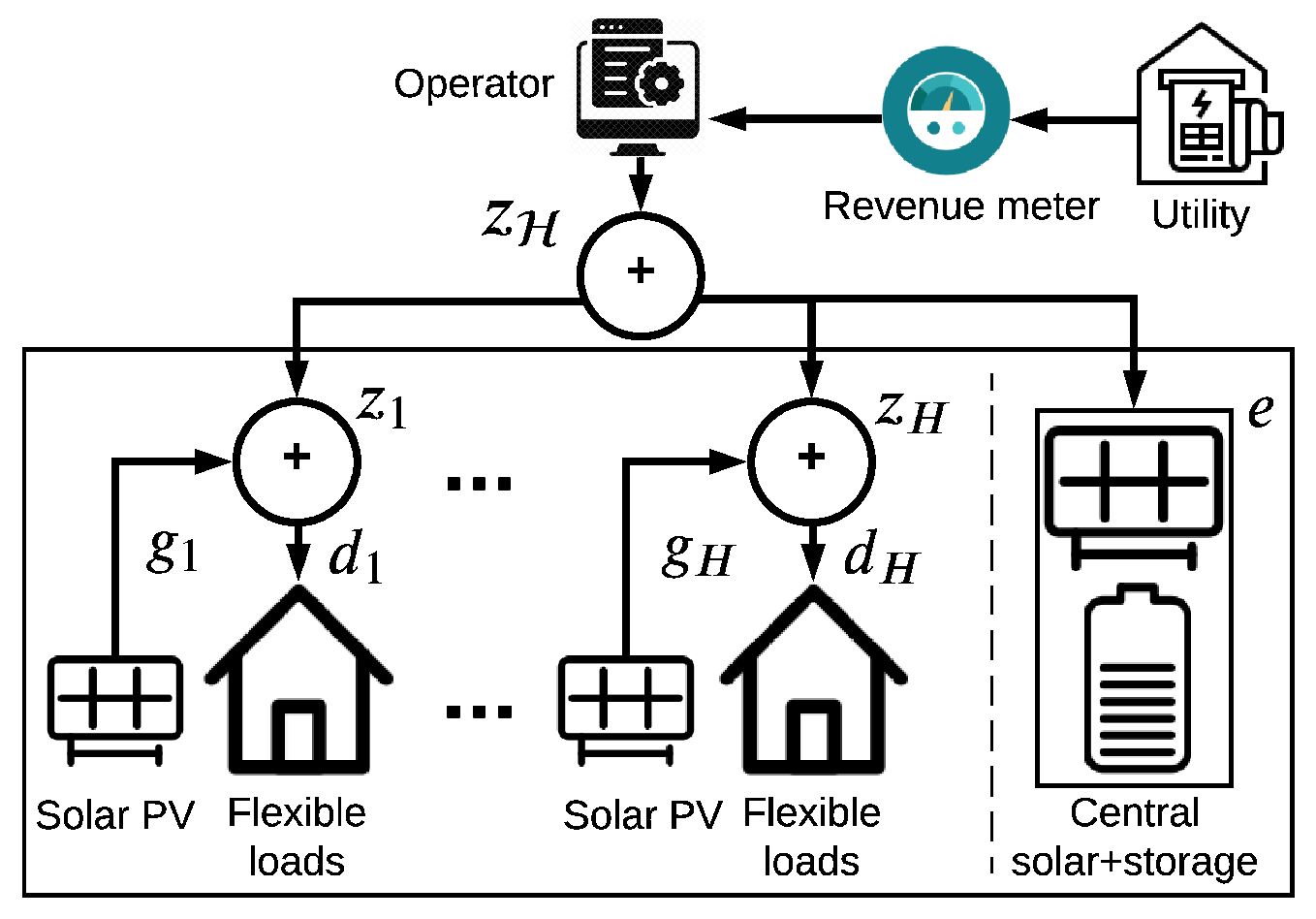}
    \caption{Energy community framework. Member consumption and renewables are $d_i, g_i \in \mathcal{R}_+$, respectively, and member net consumption, centralized resources, and aggregate net consumption are $z_i, e, z_\mathcal{H}\in \mathcal{R}$, respectively.}
    \label{fig:EnergyCommunity}
\end{figure}

\lettrine{E}{nergy} communities are regarded as a solution that improves system efficiency, economies of scale, and equity while enabling distributed energy resources (DER) aggregation and wider technology accessibility \cite{NREL_communitysolar:10NREL, Yang&Guoqiang&Spanos:21TSG,parag&sovacool:16Nature}. A generic {\em energy community} is illustrated in Fig.\ref{fig:EnergyCommunity}, where a coalition of a group of customers pool and aggregate their resources within the community and perform energy and monetary transactions with the utility company as a single entity behind a point of common coupling (PCC) downstream of the utility revenue meter \cite{parag&sovacool:16Nature}. Under the widely adopted NEM policy, the utility revenue meter measures the community's net consumption and assigns a {\em buy (retail) rate} if the community is net importing, and a {\em sell (export) rate} if the community is net exporting \cite{Alahmed&Tong:22EIRACM}. Several utilities have initiated energy-community-enabling programs, such as NEM aggregation (NEMA)\footnote{See for example, Pacific Gas and Electric company (\href{https://www.pge.com/en_US/residential/solar-and-vehicles/options/option-overview/how-to-get-started/nema/net-energy-metering-aggregation.page}{PG\&E}), California, and Baltimore Gas and Electric Company (\href{https://www.bge.com/MyAccount/MyBillUsage/Documents/Electric/Rdr_18.pdf}{BGE}), Maryland.}, for university campuses, residential complexes, and medical cities.

We focus in this work on the pricing mechanism that determines each community member's payment based on her consumption, individual-owned renewable, and her share of the community-owned DER.  We set the underlying pricing principle as maximizing community social welfare while ensuring that each member gains higher benefits than possible outside the community.  To this end, we subject the pricing mechanism to the cost-causation rule.

\subsection{Related Work}
There is a rich literature on energy communities covering optimal energy management \cite{Han&Morstyn&McCulloch:19TPS}, market mechanisms \cite{Chakraborty&Poolla&Varaiya:19TSG,Yang&Guoqiang&Spanos:21TSG}, and coordination frameworks \cite{Guerreroetal:20RSER}. Most relevant to this work is the intersection of community pricing and allocation rules \cite{Chakraborty&Poolla&Varaiya:19TSG,Yang&Guoqiang&Spanos:21TSG}, 
and optimal resource scheduling for welfare/cost optimization \cite{Fleischhacker&Auer&Lettner&Botterud:19TSG,Cui&Wang&Yan&Shi&Xiao:21TSG}.

Three energy community models have been widely discussed, each offering a different market hierarchy and flexibility to its members. The first is the decentralized model with bidirectional financial/energy transactions, i.e., peer-to-peer (P2P) transactions \cite{Morstyn&Teytelboym&Mcculloch:19TSG,Sorin&Bobo&Pinson:19TPS}. Through {\em bilateral contracts}, the P2P market structure gives full flexibility to its members to switch from being price-takers to price makers, depending on their own benefit functions. The P2P market structure is often challenged by policy and physical restrictions, data storage issues, and convergence to social optimality.

The second is the centralized model involving a community operator who schedules all resources for the benefit of the community \cite{Prete&Hobbs:16AE,Han&Morstyn&McCulloch:19TPS}. While this model has the potential to achieve the highest community overall benefits, it often comes with prohibitive computation costs and a lack of members' privacy.  Also, maximizing the total community benefits may not align with the individual benefits of its members. 

The third model, to which the work presented here belongs, is the individual scheduling of its own resources incentivized through the operator's pricing mechanism \cite{Chakraborty&Poolla&Varaiya:19TSG,Cui&Wang&Li&Xiao:20,Fleischhacker&Auer&Lettner&Botterud:19TSG,Cui&Wang&Yan&Shi&Xiao:21TSG}. The major challenge is to design a pricing mechanism that aligns rational individual decisions to community benefits in terms of achieving overall efficiency, access equity, and fairness in compensation. In \cite{Fleischhacker&Auer&Lettner&Botterud:19TSG}, a bi-level optimization of an apartment building energy community with central generation and storage was formulated to analyze pricing and energy sharing. In \cite{Cui&Wang&Yan&Shi&Xiao:21TSG}, the energy cost of a solar+storage community was minimized and then allocated based on a Nash bargaining benefit-sharing model. Although the authors ensure cooperation stability under the possibility of strategic behaviors, complying with the cost-causation principle was not considered.
The authors in \cite{Cui&Wang&Li&Xiao:20} analyze a stochastic energy community model with cost-minimizing members. An algorithm is proposed for better estimation of the stochastic game. The pricing and allocations in \cite{Fleischhacker&Auer&Lettner&Botterud:19TSG,Cui&Wang&Yan&Shi&Xiao:21TSG,Cui&Wang&Li&Xiao:20} did not consider the conformity with the cost-causation principle. 

The work of Chakraborty et al. \cite{Chakraborty&Poolla&Varaiya:19TSG} stands out as the first mechanism design under the cost-causation principle, which offers every community member a lower payment than would be outside the community. The Dynamic NEM mechanism proposed in this paper generalizes the approach in \cite{Chakraborty&Poolla&Varaiya:19TSG} to include individual surplus and community social welfare as part of the design objectives of the community pricing in a decentralized optimization framework. The consideration of community social welfare optimality necessitates designing a mechanism that not only devises payment rules as in \cite{Chakraborty&Poolla&Varaiya:19TSG}, but also set pricing rules that induce community members to achieve the welfare optimality.

To our best knowledge, Dynamic NEM proposed here is the first community energy pricing mechanism that achieves efficiency under the cost-causation principle.

\subsection{Summary of Results}
We propose Dynamic NEM—a community pricing mechanism that sets the NEM price based on available DER. Dynamic NEM uses the same prices as the utility's NEM tariff, except that the import and export prices are imposed dynamically based on the gross renewables within the community rather than individual members' net consumption and the time-of-use in the utility's NEM tariffs.

The proposed Dynamic NEM generalizes the payment rule of \cite{Chakraborty&Poolla&Varaiya:19TSG} with two significant differences.  First, Dynamic NEM prices are set {\em ex-ante} (rather than imposed {\em ex-post} in \cite{Chakraborty&Poolla&Varaiya:19TSG}) prior to elicit community members' response that achieves community social welfare maximization.   Second, Dynamic NEM induces a community-level net-zero consumption zone where the shared renewables balance the total consumption.  

We establish the following properties of Dynamic NEM: 
\begin{itemize}
    \item individual surplus maximization leads to maximum community social welfare.
\item individual surplus under Dynamic NEM is higher than the maximum surplus under utility's NEM.
\item the payment rule under Dynamic NEM satisfies cost-causation principle.
\end{itemize}
Our empirical results use real residential data to construct a hypothetical energy community, under which the benefits of community members and the grid operator are showcased.

\section{Problem Formulation}\label{sec:formulation}
To formulate the energy community, we consider a finite set of $H$ community members, indexed by $i \in \mathcal{H}=\{1,2,\ldots,H\}$, sharing their resources behind a PCC under NEM (Fig.\ref{fig:EnergyCommunity}). The members are subject to operator's pricing mechanism. We assume that community members' decision process has the same timescale as that of the NEM billing period, which allows us to adopt a single time step formulation. For billing and pricing purposes, every member's generation and net consumption are assumed to be sub-metered.
\subsection{Community Resources}
We assume each member $i\in \mathcal{H}$ has $K$ controllable devices, indexed by $k \in \mathcal{K}=\{1,2,\ldots,K\}$, whose {\em energy consumption bundle} is denoted by
\begin{equation}\label{eq:Consumption}
    \boldsymbol{d}_i=\left(d_{i 1}, \cdots, d_{i K}\right) \in \mathcal{D}_i:=\{\boldsymbol{d}_i: \underline{\boldsymbol{d}}_i \preceq \boldsymbol{d}_i \preceq \overline{\boldsymbol{d}}_i\} \subseteq \mathcal{R}_{+}^K,
\end{equation}
where $\underline{\bm{d}}_i,\overline{\bm{d}}_i$ are the consumption bundle's lower and upper limits of customer $i$, respectively. The {\em aggregate consumption} of the community is denoted by $d_{\mathcal{H}}:= \sum_{i\in \mathcal{H}} \bm{1}^\top \bm{d}_i$. 
\par Community members may own {\em renewable generation}, which we denote by $g_i \in \mathcal{R}_{+}$ for every $i\in \mathcal{H}$. The community's {\em aggregate gross generation} is $g_{\mathcal{H}}:= \sum_{i\in \mathcal{H}} g_i$. Without loss of generality, centralized solar, with every $i \in \mathcal{H}$ member owning a share $x_i \in [0,1]$, is ignored.
\par The {\em net-consumption} of every $i \in \mathcal{H}$ member $z_i \in \mathcal{R}$ and the {\em aggregate net-consumption} $z_{\mathcal{H}} \in \mathcal{R}$ are defined as
\begin{equation}\label{eq:NetCons}
    z_i:= \bm{1}^\top \bm{d}_i - g_i,~~~ z_{\mathcal{H}}:= \sum_{i \in \mathcal{H}} z_i = d_{\mathcal{H}}-g_{\mathcal{H}},
\end{equation}
where $z_i\geq 0$ ($z_{\mathcal{H}}\geq 0$) and $z_i<0$ ($z_{\mathcal{H}}<0$) represent a {\em net-consuming} and {\em net-producing} member (community), respectively.
\subsection{Community Payments}
At the revenue meter (Fig.\ref{fig:EnergyCommunity}), $z_{\mathcal{H}}$ is measured and billed based on the NEM X tariff proposed by \cite{Alahmed&Tong:22IEEETSG}. Given the NEM X tariff parameter $\pi=(\pi^+,\pi^-,\pi^0)$, {\em community payment} is
\begin{equation}\label{eq:Pcommunity}
    P_{\mathcal{H}}^\pi(z_\mathcal{H})=\pi^+ [z_{\mathcal{H}}]^++\pi^- [z_{\mathcal{H}}]^- +\pi^0,
\end{equation}
where $[x]^+:=\max\{0,x\}$ and $[x]^-:=\min\{0,x\}$ denote the positive and negative part functions for any $x \in \mathcal{R}$, respectively,  and $\pi^+, \pi^-, \pi^0 \in \mathcal{R}_+$ are the {\em retail rate}, {\em export rate}, and {\em fixed charge}, respectively. We assume $\pi^- \leq \pi^+$.

\par For every $i \in \mathcal{H}$, the payment after joining the community $P_i^{\pi_c}(\cdot)$ is determined by the payment rule with the parameter $\pi_c$. The payment before joining the community, i.e., under the utility's NEM X regime, is considered as the {\em benchmark payment} given by \cite{Alahmed&Tong:22IEEETSG} as
\begin{equation}\label{eq:Pbenchmark}
    P^\pi_{i}(z_i)=\pi^+ [z_i]^++\pi^- [z_i]^- + \pi^0/H.
\end{equation}
We assume $\pi^0$ is uniformly recovered by the $H$ members.

\subsection{Community Pricing Mechanism}
We generalize the axiomatic community pricing framework to ensure equity and efficiency \cite{Chakraborty&Poolla&Varaiya:19TSG}.  In particular, we are interested in community payment rules that satisfy the axioms of: 1) {\em individual rationality}, 2) {\em profit-neutrality}, 3) {\em equity}, 4) {\em monotonicity}, 5) {\em cost-causation penalty} and 6) {\em cost-mitigation reward}. We relegate the formal statements of the axioms to the appendix and offer instead a non-mathematical description.
{\em Individual rationality} is achieved when each community member achieves higher surplus than staying under the utility's regime. {\em Profit-neutrality} ensures that the benefit/losses of the community operator are entirely redistributed among its members, i.e. {\em budget-balance}. {\em Equity} is attained when the payments (compensations) of two community members with the same net consumption are equivalent. The {\em monotonicity} axiom ensures that having higher net consumption (net production) results in higher payment (compensation). Lastly, {\em cost-causation penalty} and {\em cost-mitigation reward} are met if members pay for causing costs and get rewarded for reducing costs, respectively. 

The following definition uses the six axioms to establish conformity with the cost causation principle \cite{Chakraborty&Poolla&Varaiya:19TSG}.
\begin{definition}[Cost-causation principle]\label{def:CausationPrinciple}
    The pricing policy and payment rules meet the {\em cost causation principle} if they satisfy axioms 1--6.
\end{definition}

\subsection{Community Surplus and Welfare Optimization}
For every $i \in \mathcal{H}$, the {\em community member surplus} $S^{\pi_c}_i(\cdot)$ and {\em benchmark surplus} $S_{i}^\pi(\cdot)$ are
\begin{equation}\label{eq:Smember}
    S^{\pi_c}_i(\cdot):= U_i(\bm{d}_i)-P_i^{\pi_c}(\cdot),~ S_{i}^\pi\left(z_i\right):=U_i\left(\bm{d}_i\right)-P_{i}^\pi\left(z_i\right),
\end{equation}
respectively, where the utility function $U_i(\bm{d}_i)$ is assumed to be additive, strictly concave, strictly increasing, and continuously differentiable with a marginal utility function $\bm{L}_i$. Therefore,
\begin{equation}\label{eq:Ldefinition}
    U_i\left(\mathbf{d}_i\right):=\sum_{k\in \mathcal{K}} U_{i k}\left(d_{ik}\right), \mathbf{L}_i:=\nabla U_i=\left(L_{i 1}, \ldots, L_{i K}\right).
\end{equation}
{\em Community welfare} is defined as the sum of community members' surpluses $W^\pi_{\mathcal{H}}:= \sum_{i \in \mathcal{H}} S^{\pi_c}_i(\cdot)$, which the operator maximizes by solving:

\begin{align} \label{eq:SurplusMax_model2}
\begin{array}{lll}\mathcal{P}_{\mathcal{H}}: &  \underset{(\bm{d}_i,\ldots,\bm{d}_H)}{\rm Maximize}& W^\pi_\mathcal{H}:=\sum_{i\in \mathcal{H}} S^{\pi_c}_i(\bm{d}_i,\cdot) \\&\text{subject to} & S^{\pi_c}_i(\bm{d}_i,\cdot)=U_i(\bm{d}_i) - P^{\pi_c}_i(\cdot),~ \forall i\\&&\sum_{i\in \mathcal{H}}  P^{\pi_c}_i(\cdot)=P^\pi_\mathcal{H}(z_\mathcal{H})\\&&z_\mathcal{H}=\sum_{i\in \mathcal{H}} (\bm{1}^\top \bm{d}_{i}-g_i)\\&&\underline{\bm{d}}_i\preceq \bm{d}_i \preceq \overline{\bm{d}}_i,~ \forall i, \end{array}  \end{align}
where the second constraint is the {\em profit-neutrality} condition.

\section{Dynamic NEM for decentralized welfare optimization}\label{sec:Models}
The community operator's primary task is to develop a pricing mechanism that induces community members to schedule their resources in a way that achieves overall welfare optimality while conforming with the cost-causation principle.
\subsection{Optimal Community Pricing}
The operator gathers every member's $i \in \mathcal{H}$ information $\mathcal{I}_i$ and uses $\mathcal{I}=\{\mathcal{I}_i,\ldots,\mathcal{I}_\mathcal{H}\}$ to solve for the optimal community strategy\footnote{We assume that the community operator learns its members' inverse marginal utility functions and consumption limits.}, which is used to envisage the pricing mechanism.

\begin{policy*}\label{alloc:ProfitNeutral}
The pricing policy for all $i\in \mathcal{H}$ members, is given by the 4-tuple parameter $\pi_c=(\pi^+,\pi^z (g_\mathcal{H}),\pi^-,\pi^{0}_c)$ with the order $\pi^+ \geq \pi^z (g_\mathcal{H}) \geq \pi^-$, where $\pi^{0}_c=\pi^0/H$, and $\pi^z(g_\mathcal{H}):=\mu^\ast(g_\mathcal{H})$ is the solution of:
 \begin{equation}\label{eq:NetZeroPrice}
 \sum_{i \in \mathcal{H}} \sum_{k \in \mathcal{K}} \max\{\underline{d}_{ik}, \min\{f_{ik}(\mu),\bar{d}_{ik}\}\} = g_\mathcal{H},
\end{equation}
where $f_{ik} := f_{ik}$ is the inverse marginal utility function for every $i \in \mathcal{H}, k \in \mathcal{K}$. The payment rule is given by
\begin{equation}\label{eq:PricingMechanism}
 P^{\pi_c}_i(z_i,g_\mathcal{H})=\begin{cases}   \pi^+ z_i + \pi^{0}_c,&\hspace{-.70em} g_\mathcal{H}<d_\mathcal{H}^+ \\ \pi^z(g_\mathcal{H})z_i + \pi^{0}_c, &\hspace{-.70em} g_\mathcal{H}\in[d_\mathcal{H}^+,d_\mathcal{H}^-] \\ \pi^- z_i + \pi^{0}_c,& \hspace{-.70em}
 g_\mathcal{H}>d_\mathcal{H}^-. \end{cases}
 \end{equation}
 where
 \begin{align}
d^+_\mathcal{H} &:=\sum_{i \in \mathcal{H}} \sum_{k \in \mathcal{K}} \max \{ \underline{d}_{ik},\min\{f_{ik}(\pi^+),\bar{d}_{ik}\}\}\label{eq:d+H}\\
d^-_\mathcal{H} &:=\sum_{i \in \mathcal{H}} \sum_{k\in \mathcal{H}} \max \{\underline{d}_{ik},\min\{f_{ik}(\pi^-),\bar{d}_{ik}\}\}\geq d^+_\mathcal{H}.\label{eq:d-H}
\end{align}
\end{policy*}
\subsubsection{Structural Properties of Dynamic NEM}
The dynamic pricing policy has an appealing threshold-based resource-aware structure, that announces community prices based on the level of aggregate renewable generation compared to the two renewable-generation-independent thresholds $d_\mathcal{H}^+$ and $d_\mathcal{H}^-$. The thresholds arise from the community's optimal aggregate consumption $d^\ast_{\mathcal{H}}$ that solves (\ref{eq:SurplusMax_model2}), given in Theorem \ref{thm:MarketEffeciency} as
\begin{equation}\label{eq:aggCons}
    d^\ast_{\mathcal{H}}(g_\mathcal{H}) = \max\{d^+_{\mathcal{H}},\min\{g_{\mathcal{H}},d^-_{\mathcal{H}}\}\}.
\end{equation}
From (\ref{eq:aggCons}), we note that $d_\mathcal{H}^+$ and $d_\mathcal{H}^-$ partition the range of $g_\mathcal{H}$ into three zones based on whether the community is 1) net-consuming ($z^\ast_\mathcal{H}>0$), 2) net-producing ($z^\ast_\mathcal{H}<0$) or 3) net-zero ($z^\ast_\mathcal{H}=0$), where $z^\ast_\mathcal{H}(g_\mathcal{H}):=d^\ast_\mathcal{H}-g_\mathcal{H}$. 

In the net-consuming and net-producing zones, the optimal community at the PCC faces the utility's $\pi^+$ and $\pi^-$, respectively, and directly passes these two prices to its members. When $g_\mathcal{H} \in [d_\mathcal{H}^+,d_\mathcal{H}^-]$, the community is energy-balanced $d^\ast_{\mathcal{H}}(g_\mathcal{H}) = g_\mathcal{H}$, and the volumetric charge is zero. It turns out that, in the net zero zone, it is optimal to charge members by the Lagrangian multiplier satisfying the Karush-Kuhn-Tucker (KKT) condition of the net zero zone -- i.e., $d_\mathcal{H}^\ast(g_\mathcal{H})=g_\mathcal{H}$. Therefore, the price $\pi^z(g_\mathcal{H})$ dynamically decreases with increasing $g_\mathcal{H}$ to incentivize demand increases, keeping the community off the grid. 

\subsubsection{Intuitions of Dynamic NEM}
The pricing policy is economically intuitive as it responds to the increasing community local generation-to-demand ratio by dynamically reducing the price from $\pi^+$ to $\pi^-$ through $\pi^z(g_\mathcal{H})$. Unlike their benchmarks in (\ref{eq:Pbenchmark}), who face the so-called {\em NEM 2.0} with different prices for imports and exports, community members under Dynamic NEM, have equivalent import and export rates, i.e., {\em NEM 1.0}.

Compared to their benchmark, net-producing community members under (\ref{eq:PricingMechanism}), are compensated at prices higher than $\pi^-$ if the community is not net-producing $g_\mathcal{H}<d_\mathcal{H}^-$. Also, net-consuming members face prices lower than $\pi^+$, if the community is not net-consuming $g_\mathcal{H}>d_\mathcal{H}^+$. This also applies to {\em non-adopting} members (i.e., customers without DG).

\subsection{Individual optimization under Dynamic NEM}
Given Dynamic NEM, every $i \in \mathcal{H}$ member maximizes its surplus (\ref{eq:Smember}) by optimally scheduling its consumption as
\begin{align} \label{eq:IndivdiualOptimization}
\mathcal{P}_i:~ \boldsymbol{d}_i^\ast~~=&~~ \underset{\boldsymbol{d}_i \in \mathcal{R}_{+}^K}{\operatorname{argmax}} \quad S_i^{\pi_c}(\cdot):=U_i\left(\boldsymbol{d}_i\right)-P^{\pi_c}_i\left(\mathbf{1}^{\top} \boldsymbol{d}_i-g_i\right) \nn \\
& \text { subject to } \quad \underline{\boldsymbol{d}}_i \preceq \boldsymbol{d}_i \preceq \overline{\boldsymbol{d}}_i.
\end{align} 



The following theorem states that, under Dynamic NEM, the aggregate optimal surplus in (\ref{eq:IndivdiualOptimization}) for all $i \in \mathcal{H}$ results in the maximum social welfare of (\ref{eq:SurplusMax_model2}).

\begin{theorem}[Decentralized welfare maximization]\label{thm:MarketEffeciency}
    Under Dynamic NEM, the community welfare maximization $\mathcal{P}_{\mathcal{H}}$ is decentrally achieved by the sum of community members' surplus maximizations $\mathcal{P}_{i}$ for all $i \in \mathcal{H}$. \hfill$\Box$
\end{theorem}

Dynamic NEM, not only induces members to achieve welfare optimality, but grants them surplus levels that are higher than their maximum surplus under the utility's NEM X $S^{\ast,\pi}_i(g_i)$ \cite{Alahmed&Tong:22EIRACM}.
\begin{theorem}[Individual rationality]\label{thm:MemberSurplus_Model2}
Under Dynamic NEM, every $i \in \mathcal{H}$ member is better off with the community, i.e., $S_i^{*, \pi_c}\left(z_i^\ast, g_{\mathcal{H}}\right) \geq S_i^{\ast, \pi}\left(g_i\right)$. \hfill$\Box$
\end{theorem}
Worth noting is that Theorem \ref{thm:MemberSurplus_Model2} applies to non-DG adopting members too, because $S^{\ast,\pi_c}_i(\bm{d}^\ast_i,g_\mathcal{H}) \geq S^{\ast,\pi}_i(0)$. As a result of Theorem \ref{thm:MemberSurplus_Model2}, the welfare of the community is higher than its benchmark of $H$ {\em optimal} standalone customers under the utility's NEM X -- i.e., $W^{\ast,\pi}(g_\mathcal{H}) \geq \sum_{i \in \mathcal{H}} S^{\ast,\pi}_i(g_i)$.

Lastly, we employ Definition \ref{def:CausationPrinciple} to show that Dynamic NEM satisfies the cost-causation principle.
\begin{theorem}[Cost-causation conformity]\label{thm:CausationPrinciple_Model2}
The payment rule under Dynamic NEM satisfies the cost-causation principle.
\hfill$\Box$
\end{theorem}

\section{Numerical Results}\label{sec:num}
To show the performance of the proposed mechanism, we assumed a hypothetical energy community of 24 households. All households have flexible loads and 19 of them have rooftop solar. We used PecanStreet data\footnote{The data is accessible at \href{http://www.pecanstreet.org/}{Pecan St. Project}.}, which has one year (2018) residential household data from Austin, TX.

To model consumption preferences, we adopted a widely-used quadratic concave utility function of the form and utility parameters learning method as given in Appendix D of \cite{Alahmed&Tong:22EIRACM}.

The community faces the utility's NEM X tariff with a time-of-use rate with $\pi^+_h=\$0.40$/kWh and $\pi^+_l=\$0.20$/kWh as peak and offpeak prices, respectively. For the export rate $\pi^-$, we used the average real-time wholesale prices\footnote{The data is accessible at: \href{https://www.ercot.com/mktinfo/prices}{ERCOT}.} in Texas in 2018. Fixed charges were assumed to be zero, i.e., $\pi^0=0$.

Two energy communities are studied and compared: 1) a community under Dynamic NEM (referred to as {\em community 1}), and 2) a community under the allocation rule in \cite{Chakraborty&Poolla&Varaiya:19TSG} (referred to as {\em community 2}), with members performing consumption decisions similar to the optimal benchmark \cite{Alahmed&Tong:22IEEETSG}.

\begin{figure}[htbp]
    \centering
    \includegraphics[scale = 0.38]{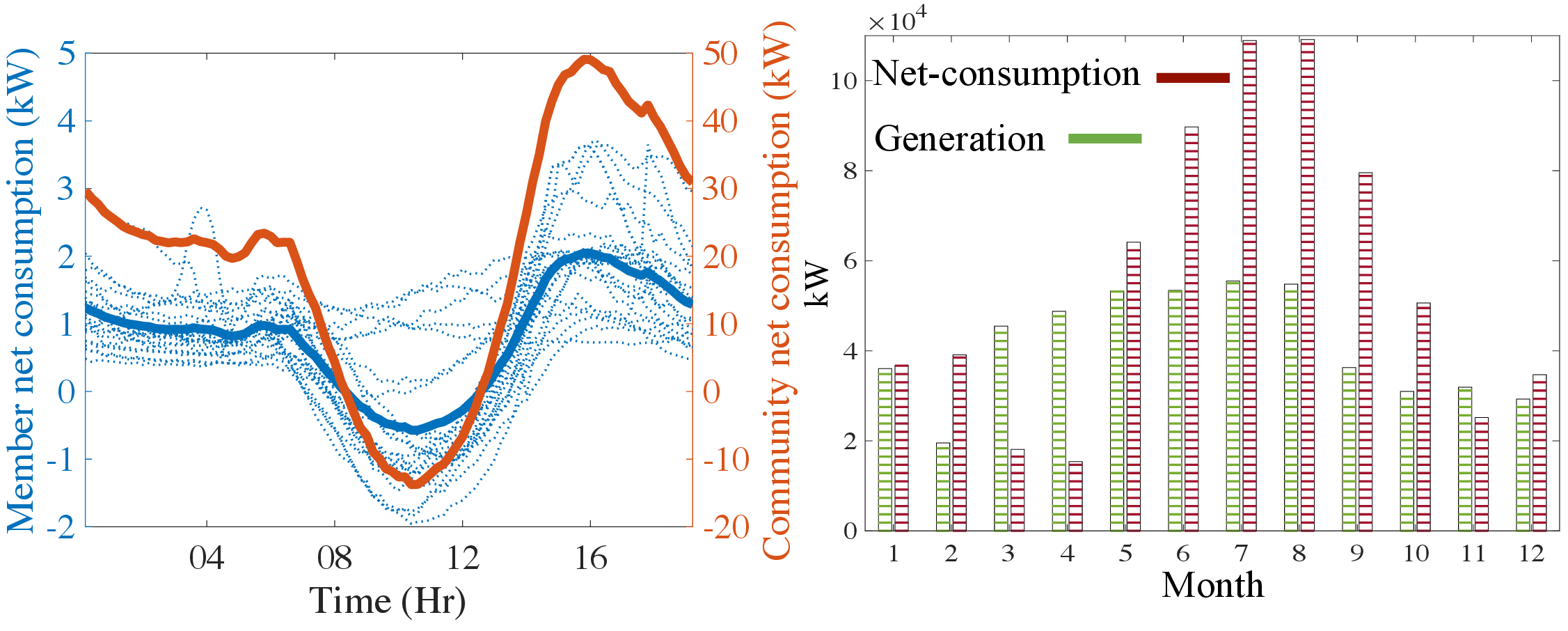}
    \caption{Left: Individual and aggregate daily net consumption. Right: Aggregate monthly net consumption and renewable generation.}
    \label{fig:rawdata}
\end{figure}

Fig.\ref{fig:rawdata} presents a summary of raw data. The left plot shows the daily average net consumption of each household (dashed blue) and their average (solid blue) in addition to the community's average net consumption (orange). During renewable generation hours, the average net consumption of many members has a different sign than the community’s average net consumption, which gives them an additional benefit as shown by Dynamic NEM. The right plot shows the community’s monthly aggregate renewable generation (green) and net consumption (red). The net consumption was much higher in summer signaling the high consumption in those months due to air-conditioning loads.

\subsection{Community Members Surplus}
\begin{figure}[htbp]
    \centering
    \includegraphics[scale = 0.37]{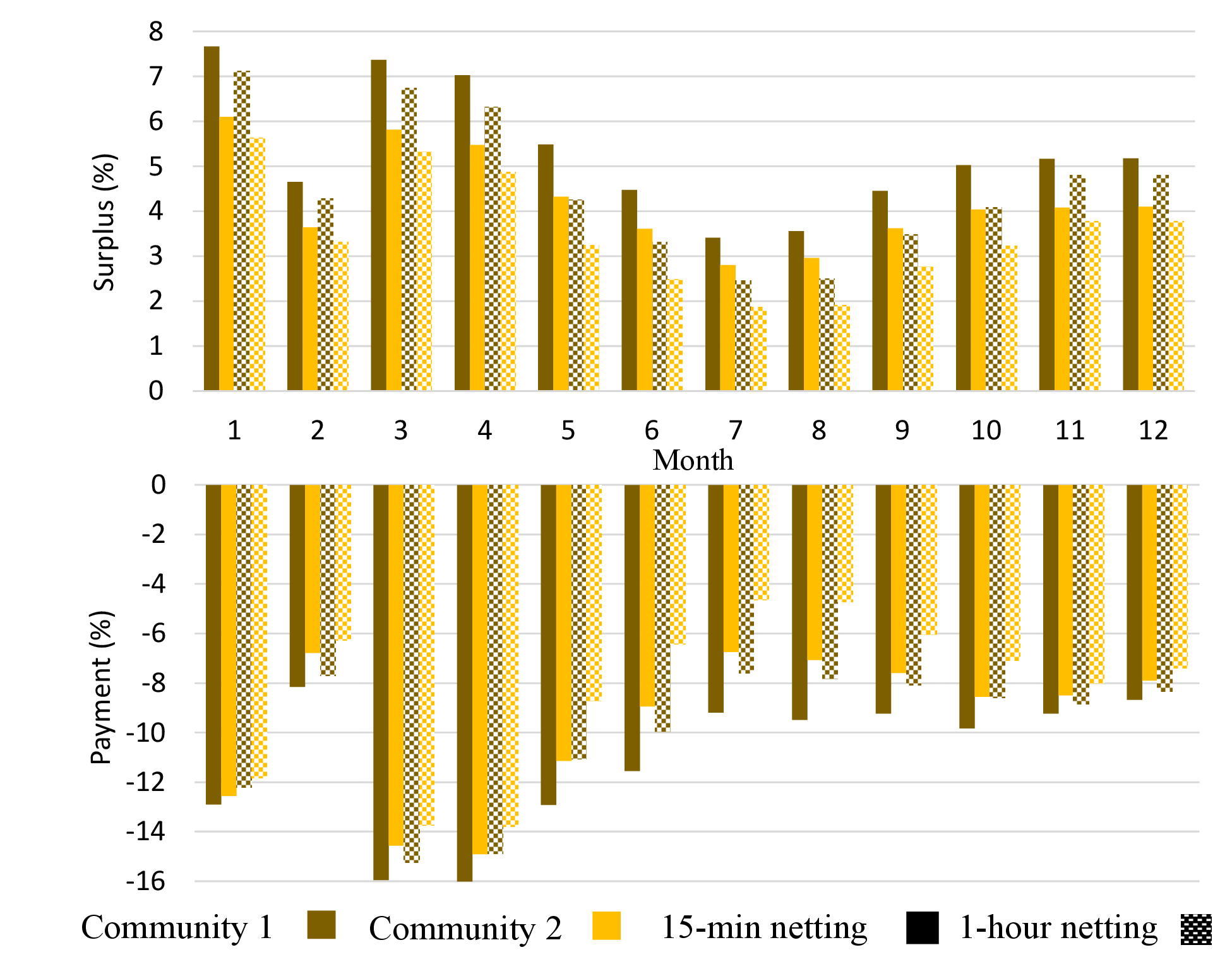}
    \caption{Community monthly surplus and payment gains (\%).}
    \label{fig:SurOpt1}
\end{figure}

Fig.\ref{fig:SurOpt1} shows the surplus (top) and payment (bottom) gains (\%) after joining communities 1 and 2 over the benchmark of optimal customers under the utility's NEM X. Two NEM net billing periods were considered; 15-minutes (solid) and 1-hour (checked). Joining either community 1 or community 2 was advantageous for the households in terms of both surplus and payments. In all months, community 1 achieved higher surpluses and lower payments than community 2. Increasing the netting frequency from hourly to 15-min increased the value of joining the communities, as the benchmark customers become more vulnerable to the export rate. Lastly, the benefit of joining the energy community was the lowest when net consumption (Fig.\ref{fig:rawdata}) was the highest, i.e., the period from June--September. This is because, in these months, community members, for most hours, face the same price their operator face at the PCC, which does not create benefits for them.

\begin{figure}[htbp]
    \centering
    \includegraphics[scale = 0.4]{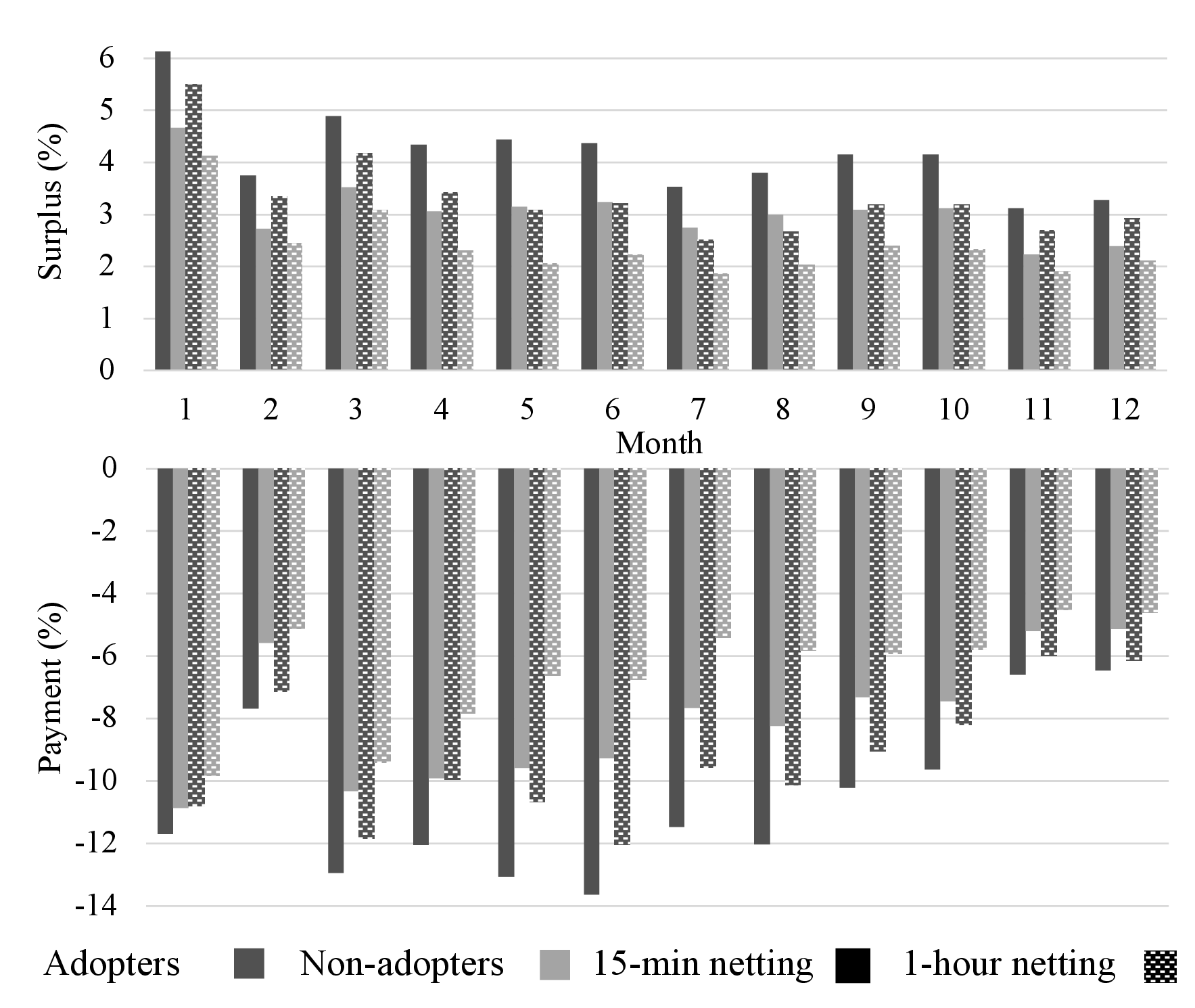}
    \caption{Adopters/non-adopters monthly surplus and payment gains (\%).}
    \label{fig:SurOpt2}
\end{figure}

Fig.\ref{fig:SurOpt2} shows the surplus (top) and payment (bottom) gains (\%) of DG adopters and non-adopters over their benchmark after joining community 1. Both classes benefited from joining the community by having higher surpluses and lower payments. However, adopters benefited more from the community as they more often operate in net consumption zones different than the community. Congruent with Fig.\ref{fig:SurOpt1}, households benefited more from the community when the netting was faster.

\subsection{Reverse Power Flows}
\begin{figure}[htbp]
    \centering
    \includegraphics[scale = 0.38]{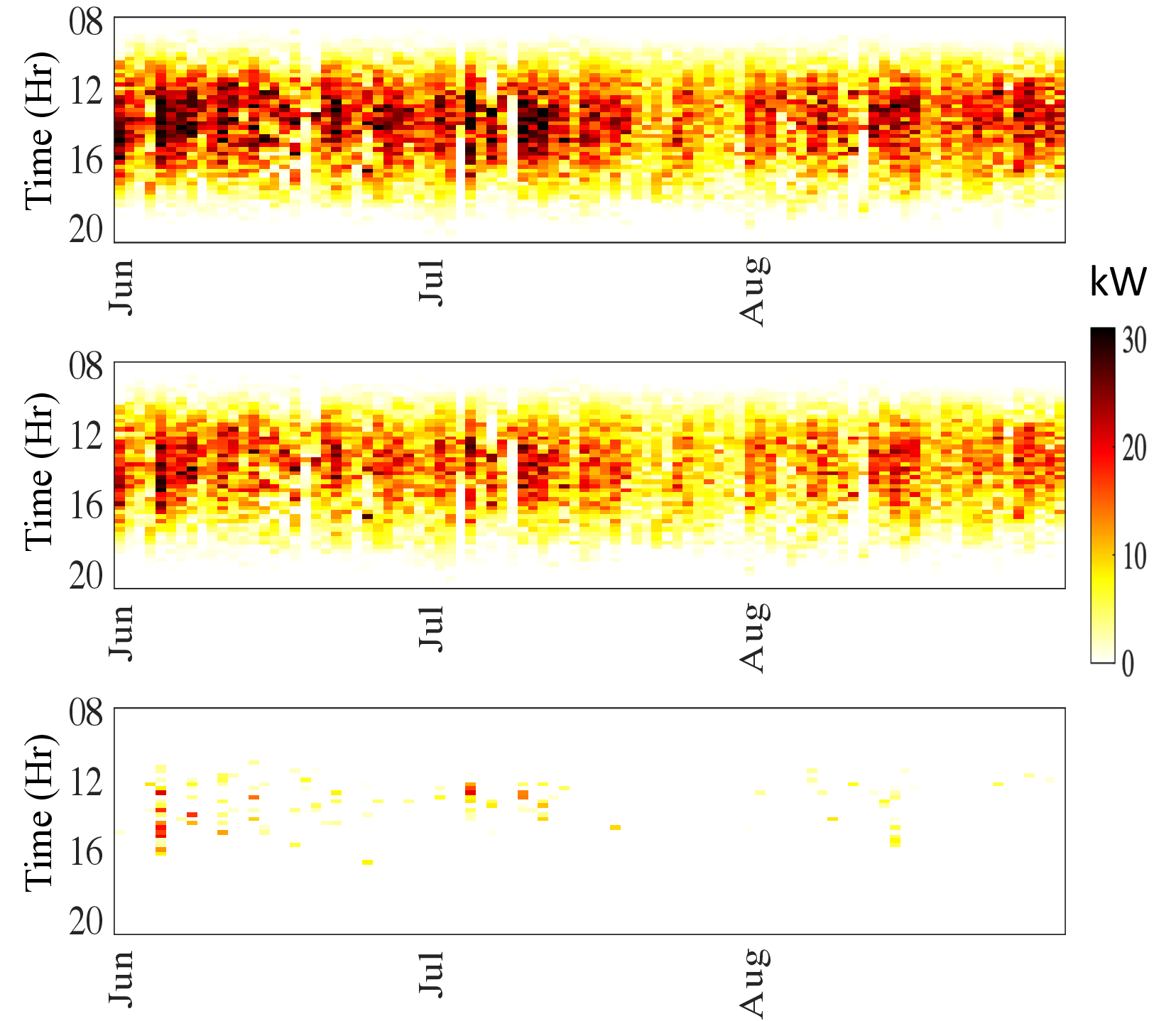}
    \caption{Aggregate RPFs (in kW) of passive utility customers (top), active utility customers (middle), and energy community (bottom).}
    \label{fig:RPF}
\end{figure}

To grid operators, energy communities can relieve network congestion and reverse power flows (RPFs)\footnote{RPFs are caused by active power injections from customer facilities to the distribution network  \cite{Sutanto_PVeffectMitigationUsingDER:13IEEETPS}.} that cause voltage instabilities \cite{Sutanto_PVeffectMitigationUsingDER:13IEEETPS}. The reduction of RPFs reduces the overall operating cost and enhances system reliability. Fig.\ref{fig:RPF} shows the aggregate RPFs of a neighborhood of passive (top) and optimal -- i.e., benchmark (middle) customers, and under community 1 (bottom) over three summer months. The benchmark customers (middle) resulted in lower RPFs than passive ones (top), as they dynamically increase their consumption to keep more of the renewables behind the utility revenue meter. The wiped-out aggregate RPFs heatmap of community 1 shows that the formation of the community diminished almost all RPF, due to sharing the renewable generation with other customers behind the PCC, which was further asserted by Dynamic NEM, which incentivized increasing the consumption when the community's renewable generation was abundant.

\section{Conclusion}\label{sec:conclusion}
 Energy communities overcome several physical, operational and financial challenges faced by standalone DER adoption. In this work, Dynamic NEM is proposed as a mechanism that decentrally achieves community welfare optimality through its surplus-maximizing members.  In addition to satisfying the cost-causation principle, the community-resource-aware Dynamic NEM attains surplus levels for its members, that are not attainable under the utility's NEM regime outside the community. The structural properties of Dynamic NEM and its economical and operational intuitions are also highlighted.
{
\bibliographystyle{IEEEtran}
\bibliography{CS_PESGM}
}

\section*{Appendix A: Cost-causation axioms}\label{sec:appendix_axioms}
For the community operator to devise justifiable and reasonable allocation rules, an axiomatic framework, containing six axioms, is used from \cite{Chakraborty&Poolla&Varaiya:19TSG}, with slight generalizations to the {\em individual rationality}, {\em cost causation penalty}, and {\em cost mitigation reward} axioms.
\begin{axiom}[Individual rationality]\label{ax:rationality}
    The community member's surplus should be at least equal to its surplus before joining the community (benchmark surplus) -- i.e., $S^{\pi_c}_i(\cdot) \geq S_{i}^\pi\left(z_i\right)$. 
\end{axiom}
For the second axiom, the following definition is needed.
\begin{definition}[Community operator profit]\label{def:profit}
The community operator’s profit $\Psi^{\pi,\pi_c}(\cdot)$ is the difference between the aggregate payments it collects from every $i \in \mathcal{H}$ member $P_i^{\pi_c}(\cdot)$, and the payment to the utility $P^\pi(z_\mathcal{H})$:
\begin{equation}\label{eq:profit}
    \Psi^{\pi,\pi_c}(\cdot):= \sum_{i \in \mathcal{H}} P^{\pi_c}_i(\cdot)-P^\pi_\mathcal{H}(z_\mathcal{H}).
\end{equation}
\end{definition}

\begin{axiom}[Profit-neutrality]\label{ax:ProfitNeutrality}
     The mechanism must ensure a profit-neutral community operator, i.e., $\Psi^{\pi,\pi_c}(\cdot)=0$.
\end{axiom}
\begin{axiom}[Equity]\label{ax:equity}
    The mechanism is {\em equitable} if, for any two community members $i,j \in \mathcal{H}, i\neq j$, having $z_i =z_j$ results in $P^{\pi_c}_i(\cdot)=P^{\pi_c}_j(\cdot)$.
\end{axiom}

\begin{axiom}[Monotonicity]\label{ax:monotonicity}
    The mechanism is {\em monotonic} if, for any two community members $i,j \in \mathcal{H}, i\neq j$, having $|z_i| \geq |z_j|$ and $z_i z_j \geq 0$ result in $|P^{\pi_c}_i(\cdot)|\geq |P^{\pi_c}_j(\cdot)|$.
\end{axiom}
For the modified axioms 5 and 6, the following generalized definition is needed \cite{Chakraborty&Poolla&Varaiya:19TSG}.
\begin{definition}[Cost causation and mitigation]\label{def:CostCausationMitigation}
    A net-consuming community member $z_i>0$ causes cost, whereas a net-producing community member $z_i <0$ mitigates cost.
\end{definition}
\par \textit{Axiom 5 (Cost causation)}: A net-consuming community member $z_i>0$ is penalized for causing cost, i.e., for any $i\in \mathcal{H}, \tilde{P}^{\pi_c}_i(\cdot)>0$ if $z_i>0$, where $\tilde{P}^{\pi_c}_i$ is the fixed-charge-adjusted $P^{\pi_c}_i$.
\par \textit{Axiom 6 (Cost mitigation)}: A net-producing community member $z_i<0$ is rewarded for mitigating cost, i.e., for any $i\in \mathcal{H}, \tilde{P}^{\pi_c}_i(\cdot)<0$ if $z_i<0$.

We note here that generalizing the individual rationality to customer surplus benchmarking rather than customer payment benchmarking as in \cite{Chakraborty&Poolla&Varaiya:19TSG} makes it stricter to satisfy the cost causation principle. In fact, under the generalized axioms, the NEM allocation rule proposed in \cite{Chakraborty&Poolla&Varaiya:19TSG}, no longer satisfy the cost causation principle.
\section*{Appendix B: Proofs}\label{sec:appendix_proofs}
\subsection{Proof of Theorem \ref{thm:MarketEffeciency}}
To prove the {\em market efficiency} of the pricing mechanism in (\ref{eq:PricingMechanism}), we solve the program $\mathcal{P}_\mathcal{H}$ in (\ref{eq:SurplusMax_model2}) and show that its optimal value is equivalent to the sum of the optimal values of the pricing-policy-dependent $\mathcal{P}_i$ in (\ref{eq:IndivdiualOptimization}) for every $i\in \mathcal{H}$ member. The solutions of $\mathcal{P}_\mathcal{H}$ and $\mathcal{P}_i$ are presented in Lemmas \ref{lem:CentralizedSoln}-\ref{lem:DecentralizedSoln}, respectively.

\begin{lemma}[Centralized welfare maximization]\label{lem:CentralizedSoln}
    The maximum social welfare, under the centralized problem $\mathcal{P}^{\mathcal{H}}$ in (\ref{eq:SurplusMax_model2}) is (assuming $\pi^0=0$):
\begin{align}
 W^{\ast,\pi}_\mathcal{H}(g_\mathcal{H})&= \begin{cases}  \sum_{i \in \mathcal{H}} U_{i}(\bm{d}^+_{i})-\pi^+ \left(d_\mathcal{H}^+ - g_\mathcal{H}^+ \right),&\hspace{-.75em} g_\mathcal{H}<d_\mathcal{H}^+ \\ \sum_{i \in \mathcal{H}}U_{i}(\bm{d}^z_{i}(g_\mathcal{H})), &\hspace{-.75em} g_\mathcal{H}\in[d_\mathcal{H}^+,d_\mathcal{H}^-] \\ \sum_{i \in \mathcal{H}} U_{i}(\bm{d}^-_{i}) - \pi^-\left(d_\mathcal{H}^- - g_\mathcal{H}^+ \right), &\hspace{-.75em} g_\mathcal{H}>d_\mathcal{H}^-, \end{cases} \label{eq:OptWelfare}
\end{align}   
where 
\begin{equation}\label{eq:d+Hd-H}
    d_\mathcal{H}^+:=\sum_{i \in \mathcal{H}} \sum_{k \in \mathcal{K}} d_{ik}^+ \quad,\quad d_\mathcal{H}^-:=\sum_{i \in \mathcal{H}} \sum_{k \in \mathcal{K}} d_{ik}^-\geq d_\mathcal{H}^+,
\end{equation}
and $\bm{d}^+_i, \bm{d}^-_i \in \mathcal{R}^K_+$ are given by
\begin{align}
\bm{d}^+_i &:= \max \{\underline{\bm{d}}_{i},\min\{\bm{f}_i(\bm{1}\pi^+),\bar{\bm{d}}_i\}\}\label{eq:d+}\\
\bm{d}^-_i &:=\max \{\underline{\bm{d}}_{i},\min\{\bm{f}_i(\bm{1}\pi^-),\bar{\bm{d}}_i\}\} \succeq \bm{d}^+_i,\label{eq:d-}
\end{align}
with the $\max,\min$ operators being element-wise and $\bm{f}_i:=\bm{L}_i^{-1}$ is the inverse marginal utility vector. The net-zero zone consumption $\bm{d}^z_i \in [\bm{d}^+_i,\bm{d}^-_i]\in \mathcal{R}^K_+$ is given by
\begin{align}
\bm{d}^z_i(g_\mathcal{H}) := \max \{\underline{\bm{d}}_{i},\min\{\bm{f}_i(\bm{1}\mu^\ast(g_\mathcal{H}),\bar{\bm{d}}_i\}\},\label{eq:do_model2}
\end{align}
where $\mu^\ast(g_\mathcal{H}) \in [\pi^-, \pi^+]$ is the Lagrangian multiplier solving:
\begin{equation}\label{eq:OptZeroZone_Model2}
 \sum_{i \in \mathcal{H}} \sum_{k \in \mathcal{K}} \max\{\underline{d}_{ik}, \min\{f_{ik}(\mu),\bar{d}_{ik}\}\} = g_\mathcal{H}.
\end{equation}
\end{lemma}
\subsubsection*{Proof of Lemma \ref{lem:CentralizedSoln}}
Note that, given {\em profit-neutrality} constraint, we can reformulate $\mathcal{P}_\mathcal{H}$ to:
\begin{align}
\begin{array}{lll}\mathcal{P}_{\mathcal{H}}: &  \underset{(\bm{d}_i,\ldots,\bm{d}_H)}{\rm maximize}&
W^{\pi}_\mathcal{H} = \sum_{i \in \mathcal{H}} U_i(\bm{d}_i) - P^\pi_\mathcal{H}(z_\mathcal{H})
\\&\text{subject to} & z_\mathcal{H}=\sum_{i\in \mathcal{H}} (\bm{1}^\top \bm{d}_{i}-g_i)\\&&\underline{\bm{d}}_i\preceq \bm{d}_i \preceq \overline{\bm{d}}_i,~ \forall i, \end{array} \nonumber  \end{align}
where $P^\pi_\mathcal{H}$ is the community payment in (\ref{eq:Pcommunity}). The program above is a slight generalization to the standalone consumer decision problem under the utility's NEM regime in \cite{Alahmed&Tong:22IEEETSG}. Therefore, the optimal decisions follow Theorem 1 in \cite{Alahmed&Tong:22IEEETSG}, but with an additional dimension representing community members.

\par From Theorem 1 in \cite{Alahmed&Tong:22IEEETSG}, and considering the additional dimension of community members, the optimal consumption $\bm{d}^{\ast}_i \in \mathcal{R}_+^K$ of every $i \in \mathcal{H}$ member becomes:
\begin{align}
    \bm{d}^{\ast}_i(g_\mathcal{H})&= \begin{cases}  \bm{d}_i^+,&\hspace{-.75em} g_\mathcal{H}<d_\mathcal{H}^+ \\ \bm{d}_i^z(g_\mathcal{H}), &\hspace{-.75em} g_\mathcal{H}\in[d_\mathcal{H}^+,d_\mathcal{H}^-] \\ \bm{d}_i^-, &\hspace{-.75em} g_\mathcal{H}>d_\mathcal{H}^-,\end{cases} \label{eq:Optmemberd} 
\end{align}
where $\bm{d}^+_\mathcal{H}, \bm{d}^-_\mathcal{H}$ are as defined in (\ref{eq:d+Hd-H}), and $\bm{d}_i^+, \bm{d}_i^z, \bm{d}_i^-$ are as defined in (\ref{eq:d+})-(\ref{eq:do_model2}). The payment of the community at the PCC under the optimal consumption decisions in (\ref{eq:Optmemberd}) becomes a function of $g_\mathcal{H}$ as (assuming $\pi^0=0$):
\begin{align}
    P^{\ast,\pi}_\mathcal{H}(z^\ast_i,g_\mathcal{H})&=  \begin{cases}  \pi^+(d_\mathcal{H}^+-g_\mathcal{H}),&\hspace{-.75em} g_\mathcal{H}<d_\mathcal{H}^+ \\ 0, &\hspace{-.75em} g_\mathcal{H}\in[d_\mathcal{H}^+,d_\mathcal{H}^-] \\ \pi^-(d_\mathcal{H}^--g_\mathcal{H}), &\hspace{-.75em} g_\mathcal{H}>d_\mathcal{H}^-.\end{cases} \label{eq:OptCommunityP} 
\end{align}
The social welfare under optimal decisions is
\begin{align}
W^{\ast,\pi}_{\mathcal{H}} &= \sum_{i\in \mathcal{H}} S_i^{\pi_c}(\cdot)= \sum_{i\in \mathcal{H}} U_i(\bm{d}^\ast_i) -P_i^{\pi_c}(\cdot)\nn\\ &= \sum_{i\in \mathcal{H}} U_i(\bm{d}^\ast_i) -P^{\ast,\pi}_\mathcal{H}(z^\ast_i,g_\mathcal{H}).\nn
\end{align}
By substituting (\ref{eq:Optmemberd}) and (\ref{eq:OptCommunityP}) above, one should get the maximum social welfare in (\ref{eq:OptWelfare}). \QED

\begin{lemma}[Individual surplus maximization]\label{lem:DecentralizedSoln}
    Under Dynamic NEM, the optimal surplus for every $i \in \mathcal{H}$ member is (assuming $\pi^0=0$):
    \begin{equation} \label{eq:OptmemberS}
        S^{\ast,\pi_c}_i(z^\ast_i,g_\mathcal{H}) = \begin{cases}  U_i(\bm{d}^+_i)-\pi^+ z^\ast_i(g_\mathcal{H}),&\hspace{-.75em} g_\mathcal{H}<d_\mathcal{H}^+ \\ U_i(\bm{d}^z_i)-\pi^z(g_\mathcal{H}) z^\ast_i(g_\mathcal{H}), &\hspace{-.75em} g_\mathcal{H}\in[d_\mathcal{H}^+,d_\mathcal{H}^-] \\ U_i(\bm{d}^-_i)-\pi^- z^\ast_i(g_\mathcal{H}), &\hspace{-.75em} g_\mathcal{H}>d_\mathcal{H}^-,\end{cases}  
    \end{equation}
    where the optimal net consumption $z^\ast_i$ for every $i \in \mathcal{H}$ member is given by
    \begin{equation} \label{eq:Optmemberz}
        z^\ast_i(g_\mathcal{H}) = \begin{cases}  \sum_{i\in \mathcal{K}} d^+_{ik} - g_i,&\hspace{-.75em} g_\mathcal{H}<d_\mathcal{H}^+ \\ \sum_{i\in \mathcal{K}} d^z_{ik} - g_i, &\hspace{-.75em} g_\mathcal{H}\in[d_\mathcal{H}^+,d_\mathcal{H}^-] \\ \sum_{i\in \mathcal{K}} d^-_{ik} - g_i, &\hspace{-.75em} g_\mathcal{H}>d_\mathcal{H}^-.\end{cases} 
    \end{equation}
\end{lemma}

\subsubsection*{Proof of Lemma \ref{lem:DecentralizedSoln}}
Under Dynamic NEM, the community member surplus maximization problem in (\ref{eq:IndivdiualOptimization}) is a convex optimization problem with linear constraints.
When $g_\mathcal{H}<d^+_\mathcal{H}$, the announced price is $\pi^+$ and using the KKT conditions, every $i\in \mathcal{H}$ member's optimal consumption is $\bm{d}^\ast_i=\bm{d}^+_i$, computed as in (\ref{eq:d+}). Similarly, if $g_\mathcal{H}>d^-_\mathcal{H}$, the announced price is $\pi^-$ and every $i\in \mathcal{H}$ member's optimal consumption is $\bm{d}^\ast_i=\bm{d}^-_i$, computed as in (\ref{eq:d-}). In the net-zero zone, when $g_\mathcal{H} \in [d^+_\mathcal{H},d^-_\mathcal{H}]$, the announced price is $\mu^\ast(g_\mathcal{H})$ and every $i\in \mathcal{H}$ member consumes $\bm{d}^\ast_i=\bm{d}^z_i$. 
\par The optimal net consumption $z^\ast_i(g_\mathcal{H})$ of every $i\in \mathcal{H}$ member, therefore, is equivalent to (\ref{eq:Optmemberz}), and the payment is (assuming $\pi^0=0$):
\begin{align}
    P^{\ast,\pi_c}_i(z^\ast_i,g_\mathcal{H})&=  \begin{cases}  \pi^+z^\ast_i(g_\mathcal{H}),&\hspace{-.75em} g_\mathcal{H}<d_\mathcal{H}^+ \\ \pi^z(g_\mathcal{H}) z^\ast_i(g_\mathcal{H}), &\hspace{-.75em} g_\mathcal{H}\in[d_\mathcal{H}^+,d_\mathcal{H}^-] \\ \pi^- z^\ast_i(g_\mathcal{H}), &\hspace{-.75em} g_\mathcal{H}>d_\mathcal{H}^-.\end{cases} \label{eq:OptMemberP} 
\end{align}
By substituting the optimal member decisions $\bm{d}^\ast_i$ and payment under optimal decisions in (\ref{eq:OptMemberP}) into the community member surplus expression in (\ref{eq:Smember}), one should get (\ref{eq:OptmemberS}). \QED

Now it remains to show that $\sum_{i}^H S^{\ast,\pi_c}_i(z^\ast_i,g_\mathcal{H}) = W^{\ast,\pi}_\mathcal{H}(g_\mathcal{H})$ to prove Theorem \ref{thm:MarketEffeciency}. Note that:
\begin{equation*}
        \sum_{i}^H z^\ast_i(g_\mathcal{H}) = \begin{cases}  d^+_\mathcal{H} - g_\mathcal{H},&\hspace{-.75em} g_\mathcal{H}<d_\mathcal{H}^+ \\ 0, &\hspace{-.75em} g_\mathcal{H}\in[d_\mathcal{H}^+,d_\mathcal{H}^-] \\ d^-_\mathcal{H} - g_\mathcal{H}, &\hspace{-.75em} g_\mathcal{H}>d_\mathcal{H}^-,\end{cases} 
    \end{equation*}
where we used (\ref{eq:do_model2}), to show that $z^\ast_i(g_\mathcal{H})=0$ when $g_\mathcal{H}\in[d_\mathcal{H}^+,d_\mathcal{H}^-]$. Taking the sum over (\ref{eq:OptmemberS}), we get:
\begin{align}
       \sum_{i}^H &S^{\ast,\pi_c}_i(z^\ast_i,g_\mathcal{H})\nn \\&= \begin{cases}  \sum_{i}^H U_i(\bm{d}^+_i)-\pi^+ \left(d^+_\mathcal{H} - g_\mathcal{H}\right),&\hspace{-.75em} g_\mathcal{H}<d_\mathcal{H}^+ \\ \sum_{i}^H U_i(\bm{d}^z_i), &\hspace{-.75em} g_\mathcal{H}\in[d_\mathcal{H}^+,d_\mathcal{H}^-] \\ \sum_{i}^H U_i(\bm{d}^-_i)-\pi^- \left(d^-_\mathcal{H} - g_\mathcal{H}\right), &\hspace{-.75em} g_\mathcal{H}>d_\mathcal{H}^-,\end{cases}\nn\\&= W^{\ast,\pi}_\mathcal{H}(g_\mathcal{H}),\nn 
    \end{align}
    which proves the {\em market efficiency} under Dynamic NEM. \QED
\subsection{Proof of Theorem \ref{thm:MemberSurplus_Model2}}
We recall the active customer's surplus under the utility's NEM X regime, derived in \cite{Alahmed&Tong:22IEEETSG, Alahmed&Tong:22EIRACM} as:
\begin{align}\label{eq:explicitSbench}
    S&^{\ast,\pi}_i(g_i)=\nn \\& \begin{cases}   U_{i}(\bm{d}^+_i)-\pi^+ (\bm{1}^\top \bm{d}^+_i-g_i),&\hspace{-.75em} g_i< \bm{1}^\top \bm{d}^+_i \\ U_{i}(\bm{d}^o_i(g_i)),&\hspace{-.75em} g_i\in [\bm{1}^\top \bm{d}^+_i,\bm{1}^\top \bm{d}^-_i] \\ U_{i}(\bm{d}^-_i)-\pi^- (\bm{1}^\top \bm{d}^-_i-g_i),&\hspace{-.75em} g_i> \bm{1}^\top \bm{d}^-_i,  \end{cases}
\end{align}
where $\bm{d}^o_i(g_i)$ is customer's $i$ consumption in the net-zero zone.
Under Dynamic NEM, the surplus of every $i \in \mathcal{H}$ community member is given by (from Lemma \ref{lem:DecentralizedSoln}):
\begin{align}\label{eq:explicitScomm}
    &S^{\ast,\pi_c}_i(z^\ast_i,g_\mathcal{H})=\nn\\& \begin{cases}   U_i(\bm{d}_{i}^+) -\pi^+(\bm{1}^\top \bm{d}_{i}^+-g_i),&\hspace{-.70em} g_\mathcal{H}<d_\mathcal{H}^+ \\ U_i(\bm{d}_i^z(g_\mathcal{H}))-\pi^z(g_\mathcal{H})(\bm{1}^\top \bm{d}_i^z(g_\mathcal{H})-g_i), &\hspace{-.70em} g_\mathcal{H}\in[d_\mathcal{H}^+,d_\mathcal{H}^-] \\ U_i(\bm{d}_{i}^-)- \pi^-(\bm{1}^\top \bm{d}_{i}^- -g_i),& \hspace{-.70em}
 g_\mathcal{H}>d_\mathcal{H}^-, \end{cases}
\end{align}
where in both (\ref{eq:explicitSbench}--\ref{eq:explicitScomm}), we assumed $\pi^0=0$ for brevity. Additionally, we assume, without loss of generalit,y and for notational brevity, the single device case $K=1$, and non-binding consumption upper and lower limits.

For every customer $i \in \mathcal{H}$, the value of joining the community is $S^{\ast,\pi_c}_i(z^\ast_i,g_\mathcal{H})- S^{\ast,\pi}_i(g_i)$, which expands to
\begin{align}
&S^{\ast,\pi_c}_i(z_i^\ast,g_\mathcal{H})- S^{\ast,\pi}_i(g_i)=\nn\\&
\small{\begin{cases}0, &\hspace{-.75em} g_\mathcal{H}<d_\mathcal{H}^+, g_i<d^+_i \\ U_i(d^+_i)-U_i(g_i)-\pi^+\tilde{d}^+_i , &\hspace{-.75em} g_\mathcal{H}<d_\mathcal{H}^+, g_i \in [d^+_i,d^-_i] \\ U_i(d^+_i)-U_i(d_i^-)-\pi^+\tilde{d}^+_i+\pi^-\tilde{d}^-_i , &\hspace{-.75em} g_\mathcal{H}<d_\mathcal{H}^+, g_i >d^-_i \\ U_i(d_i^z)-U_i(d^+_i)-\pi^z\tilde{d}^z_i+\pi^+\tilde{d}^+_i , &\hspace{-.75em} g_\mathcal{H}\in [d_\mathcal{H}^+,d_\mathcal{H}^-], g_i<d^+_i \\ U_i(d_i^z)-U_i(g_i)-\pi^z\tilde{d}^z_i , &\hspace{-.75em} g_\mathcal{H}\in [d_\mathcal{H}^+,d_\mathcal{H}^-], g_i\in [d^+_i,d^-_i] \\ U_i(d_i^z)-U_i(d^-_i)-\pi^z\tilde{d}^z_i+\pi^-\tilde{d}^-_i , &\hspace{-.75em} g_\mathcal{H}\in [d_\mathcal{H}^+,d_\mathcal{H}^-], g_i>d^-_i \\ U_i(d^-_i)-U_i(d_i^+)-\pi^-\tilde{d}^-_i+\pi^+\tilde{d}^+_i , &\hspace{-.75em} g_\mathcal{H}>d_\mathcal{H}^-, g_i <d^+_i \\ U_i(d^-_i)-U_i(g_i)-\pi^-\tilde{d}^-_i , &\hspace{-.75em} g_\mathcal{H}>d_\mathcal{H}^-, g_i\in [d^+_i,d^-_i]  \\ 0 , &\hspace{-.75em} g_\mathcal{H}>d_\mathcal{H}^-, g_i> d^-_i,\end{cases}}\label{eq:DeltaSproof}
\end{align}
where $\tilde{d}^+_i:=d^+_i-g_i, \tilde{d}^z_i:=d^z_i-g_i$ and $\tilde{d}^-_i:=d^-_i-g_i$ are the DG-adjusted consumptions in the community's net-consumption, net-zero and net-production zones, respectively.

Given our assumptions on the utility function $U(\cdot)$, we use the following standard property of concave and continuously differentiable functions (also known as the {\em Rooftop Theorem}),
\bea
U(y) &\leq& U(x)+L(x)(y-x)\nn\\
U(x) &\leq& U(y)+L(y)(x-y),\nn
\eea

with $x<y,\forall x,y \in \mathcal{R}$, and where $L(\cdot)$ is the marginal utility function defined in (\ref{eq:Ldefinition}). The second inequality can be re-written as $L(y)(y-x) \leq U(y)-U(x)$. Therefore, it follows that,
\beq \label{eq:ConcavityPropoerty}
L(x)\geq \frac{U(y)-U(x)}{y-x}\geq L(y).
\eeq

We utilize the above properties to show the non-negativity of the surplus difference, by proving the non-negativity of every piece in (\ref{eq:DeltaSproof}).

--  \underline{Pieces 1 and 9}: In piece 1, the consumption of the community member and its benchmark is $d^+_i$, which is a function of $\pi^+$ that is the same before and after joining the community. The same goes for piece 9, where both the community member and its benchmark consume $d^-_i$.

--  \underline{Piece 2}: Given that $g_i>d_i^+$, we have:
\begin{align*}
\begin{gathered}
L_i\left(d_i^{+}\right) \geq \frac{U_i\left(g_i\right)-U_i\left(d_i^{+}\right)}{g_i-d_i^{+}} \geq L_i\left(g_i\right) \\
\pi^{+}\left(d_i^{+}-g_i\right) \leq U_i\left(d_i^{+}\right)-U_i\left(g_i\right) \leq \pi^z\left(d_i^{+}-g_i\right) \\
0 \leq U_i\left(d_i^{+}\right)-U_i\left(g_i\right)-\pi^{+}\left(d_i^{+}-g_i\right) \leq\left(\pi^z-\pi^{+}\right)\left(d_i^{+}-g_i\right)
\end{gathered}
\end{align*}
where we used $L_i(d^+_i)=\pi^+$. Therefore, the surplus difference in piece 2 is non-negative.

\par --  \underline{Piece 3}: Given that $d_i^->d_i^+$, we have:
\begin{equation*}
\pi^+(d^+_i-d^-_i)\leq U_i(d^+_i)-U_i(d^-_i)\leq \pi^- (d^+_i-d^-_i).
\end{equation*}

Using the lower bound $\pi^+(d^+_i-d^-_i)$, the third piece becomes $(\pi^+-\pi^-)(g_i-d^-_i)\geq0$, because $g_i> d^-_i$ and $\pi^+\geq \pi^-$.

--  \underline{Piece 4}: Given that $d_i^+<d_i^z(g_\mathcal{H})$ (see Lemma \ref{lem:CentralizedSoln}), we have:
\begin{equation*}
\pi^+ (d_i^z(g_\mathcal{H})-d^+_i)\geq U_i(d_i^z(g_\mathcal{H}))-U_i(d^+_i)\geq \pi^z (d_i^z(g_\mathcal{H})-d^+_i).
\end{equation*}

Using the lower bound $\pi^z (d_i^z(g_\mathcal{H})-d^+_i)$, we get $(\pi^+-\pi^z)(d^+_i-g_i)\geq0$, because $\pi^+\leq \pi^z$ and $g_i<d^+_i$.

-- \underline{Piece 5}: Here, we have two cases.
\par Case 1: If $d_i^z(g_\mathcal{H})> g_i$, we have:
\begin{equation*}
L(g) (d_i^z(g_\mathcal{H})-g_i)\geq U_i(d_i^z(g_\mathcal{H}))-U_i(g_i)\geq \pi^z (d_i^z(g_\mathcal{H})-g_i).
\end{equation*}
Using the lower bound $\pi^z (d_i^z(g_\mathcal{H})-g_i)$, we get $0$.
\par Case 2: If $d_i^z(g_\mathcal{H})< g_i$, then
\begin{equation*}
\pi^z (g_i-d_i^z(g_\mathcal{H}))\geq U_i(g_i)-U_i(d_i^z(g_\mathcal{H}))\geq L_i(g_i) (g_i-d_i^z(g_\mathcal{H})).
\end{equation*}
Multiplying by (-1) and using the lower bound $\pi^z (d_i^z(g_\mathcal{H})-g_i)$ we get $0$.

-- \underline{Piece 6}: Given that $d_i^z(g_\mathcal{H})<d^-_i$, we have:
\begin{equation*}
\pi^z (d_i^z(g_\mathcal{H})-d^-_i)\leq U_i(d_i^z(g_\mathcal{H}))-U_i(d^-_i)\leq \pi^- (d_i^z(g_\mathcal{H})-d^-_i).
\end{equation*}

Using the lower bound $\pi^z (d_i^z(g_\mathcal{H})-d^-_i)$, we get $(\pi^z-\pi^-)(g_i-d^-_i)\geq0$, because $\pi^z\geq\pi^-$ and $g_i>d^-_i$.

--  \underline{Piece 7}: Given $d^-_i>d^+_i$, we have:
\begin{equation*}
\pi^+(d^-_i-d^+_i)\geq U_i(d^-_i)-U_i(d^+_i)\geq \pi^- (d^-_i-d^+_i).
\end{equation*}
Using the lower bound $\pi^- (d^-_i-d^+_i)$, we get $(\pi^+-\pi^-)(d_i^+-g_i)\geq0$, because $\pi^+\geq \pi^-$ and $d_i^+>g_i$.

--  \underline{Piece 8}: Given $d^-_i>g_i$, we have:
\begin{equation*}
    L_i(g_i)(d^-_i-g_i)\geq U_i(d^-_i)-U_i(g_i)\geq \pi^- (d^-_i-g_i).
\end{equation*}
The inequality $U_i(d^-_i)-U_i(g_i)\geq \pi^- (d^-_i-g_i)\geq 0$ proves that piece 8 is non negative, because $d_i>g_i$.\\
By analyzing all 9 cases, we proved that $S^{\ast,\pi_c}_i(z_i^\ast,g_\mathcal{H})- S^{\ast,\pi}_i(g_i) \geq 0, \forall i\in \mathcal{H}$. \QED

\subsection{Proof of Theorem \ref{thm:CausationPrinciple_Model2}}
According to Def.\ref{def:CausationPrinciple}, to show the cost-causation conformity, we need to show that Dynamic NEM satisfies the six axioms.
\begin{enumerate}[labelindent=0pt]
    \item Individual rationality: From Theorem \ref{thm:MemberSurplus_Model2}, we have $S^{\ast,\pi_c}_i(z^\ast_i,g_\mathcal{H})\geq S^{\ast,\pi}_i(g_i)$, which proves the individual rationality axiom.
    \item Profit-neutrality: Using Definition \ref{def:profit}, we can show profit-neutrality by comparing the aggregated community member payments under the pricing policy $(\sum_{i\in \mathcal{H}} P^{\ast,\pi_c}_i(z^\ast_i,g_\mathcal{H}))$ to the payment of the community operator to the utility. Under optimal decisions, the payment of the community operator to the utility is given in Lemma \ref{lem:CentralizedSoln} as
    \begin{equation*}
    P^{\ast,\pi}_\mathcal{H}(g_\mathcal{H})=\pi^0 + \begin{cases} \pi^+(d_\mathcal{H}^+-g_\mathcal{H}),&\hspace{-.75em} g_\mathcal{H}<d_\mathcal{H}^+ \\ 0,&\hspace{-.75em} g_\mathcal{H}\in[d_\mathcal{H}^+,d_\mathcal{H}^-] \\ \pi^-(d_\mathcal{H}^--g_\mathcal{H}),&\hspace{-.75em} g_\mathcal{H}>d_\mathcal{H}^-, \end{cases}
\end{equation*}
and the payment of every $i \in \mathcal{H}$ community member to the operator is also given in Lemma \ref{lem:DecentralizedSoln} as
\begin{equation*}
    P^{\ast,\pi_c}_i(z^\ast_i, g_\mathcal{H})=  \pi^{0}_c+ \begin{cases}  \pi^+z^\ast_i(g_\mathcal{H}),&\hspace{-.75em} g_\mathcal{H}<d_\mathcal{H}^+ \\ \pi^z(g_\mathcal{H}) z^\ast_i(g_\mathcal{H}), &\hspace{-.75em} g_\mathcal{H}\in[d_\mathcal{H}^+,d_\mathcal{H}^-] \\ \pi^- z^\ast_i(g_\mathcal{H}), &\hspace{-.75em} g_\mathcal{H}>d_\mathcal{H}^-.\end{cases} 
\end{equation*}
    Given $\pi^{0}_c=\pi^0/H$ and, from Theorem \ref{thm:MarketEffeciency}, 
    \begin{equation*}
        \sum_{i}^H z^\ast_i(g_\mathcal{H}) = \begin{cases}  d^+_\mathcal{H} - g_\mathcal{H},& g_\mathcal{H}<d_\mathcal{H}^+ \\ 0, & g_\mathcal{H}\in[d_\mathcal{H}^+,d_\mathcal{H}^-] \\ d^-_\mathcal{H} - g_\mathcal{H}, & g_\mathcal{H}>d_\mathcal{H}^-,\end{cases} 
    \end{equation*}
one should easily see that $\Psi^{\pi,\pi_c}(g_\mathcal{H})=0$.
    \item Equity: Because the fixed charges and prices that members face in each zone is uniform, it holds that if two members $i,j \in \mathcal{H}, i \neq j$ have the same optimal net-consumption $z_i^\ast (g_\mathcal{H}) = z_j^\ast (g_\mathcal{H})$, their payments under Dynamic NEM are identical $P^{\ast,\pi_c}_i(z^\ast_i, g_\mathcal{H})=P^{\ast,\pi_c}_j(z^\ast_j,g_\mathcal{H})$. Hence, Dynamic NEM equity is shown.
    \item Monotonicity: Under Dynamic NEM, if two members $i,j \in \mathcal{H}, i \neq j$ have $z_i^\ast (g_\mathcal{H}) z_j^\ast (g_\mathcal{H}) \geq 0$ and $|z_i^\ast (g_\mathcal{H})| \geq |z_j^\ast (g_\mathcal{H})|$, then $|P^{\ast,\pi_c}_i(z^\ast_i,g_\mathcal{H})| \geq |P^{\ast,\pi_c}_j(z^\ast_j,g_\mathcal{H})|$. Hence, monotonicity of Dynamic NEM is shown.
    \item Cost-causation penalty: Under Dynamic NEM, any cost-causing member $i\in \mathcal{H}$, i.e., positive net-consumption, will have a non-negative volumetric charge submitted to the operator, no matter what the community's zone is. That is, for any $i \in \mathcal{H}$, when $z^\ast_i (g_\mathcal{H})>0$, we have $\tilde{P}^{\ast,\pi_c}_i(z^\ast_i,g_\mathcal{H})>0$. 
    \item Cost-mitigation reward: Under Dynamic NEM, any cost-mitigating member $i\in \mathcal{H}$, i.e., negative net-consumption, will get rewarded by the operator, no matter what the community's zone is. That is, for any $i \in \mathcal{H}$, when $z^\ast_i (g_\mathcal{H})<0$, we have $\tilde{P}^{\ast,\pi_c}_i(z^\ast_i,g_\mathcal{H})<0$.  \QED
\end{enumerate}

\end{document}